\documentclass[12pt,a4paper]{article}
\usepackage{graphicx}
\usepackage{times}
\title{\bf Light Curve Parameters of Cepheid and RR Lyrae Variables at Multiple Wavelengths $-$ Models vs. Observations}

\author{Harinder P. Singh$^1$\thanks{hpsingh@physics.du.ac.in} , Susmita Das$^1$, Anupam Bhardwaj$^2$, Shashi Kanbur$^3$, \\Marcella Marconi$^4$\\
\vspace{0.5cm}\\
\normalsize $^1$ Department of Physics \& Astrophysics, University of Delhi, Delhi 110007, India \\
\normalsize $^2$ Kavli Institute for Astronomy and Astrophysics, Peking University, Beijing 100871, China\\
\normalsize $^3$ State University of New York, Oswego, NY 13126, USA\\
\normalsize $^4$ INAF-Osservatorio astronomico di Capodimonte, Via Moiariello 16, 80131 Napoli, Italy}
\date{\mbox{}}
\begin{document}
\maketitle
\setcounter{page}{1001}
\pagestyle{plain}
    \makeatletter
    \renewcommand*{\pagenumbering}[1]{%
       \gdef\thepage{\csname @#1\endcsname\c@page}%
    }
    \makeatother
\pagenumbering{arabic}

%
%
\def\bull{\vrule height .9ex width .8ex depth -.1ex}
\makeatletter
\def\ps@plain{\let\@mkboth\gobbletwo
\def\@oddhead{}\def\@oddfoot{\hfil\scriptsize\bull\quad
"2nd Belgo-Indian Network for Astronomy \& astrophysics (BINA) workshop'', held in Brussels (Belgium), 9-12 October 2018 \quad\bull}%
\def\@evenhead{}\let\@evenfoot\@oddfoot}
\makeatother
%
%
\def\beginrefer{\section*{References}%
\begin{quotation}\mbox{}\par}
\def\refer#1\par{{\setlength{\parindent}{-\leftmargin}\indent#1\par}}
\def\endrefer{\end{quotation}}
%
%
{\noindent\small{\bf Abstract:} 
We present results from a comparative study of light curves of Cepheid and RR Lyrae stars in
the Galaxy and the Magellanic Clouds with their theoretical models generated from the stellar pulsation codes.
Fourier decomposition method is used to analyse the theoretical and the observed light curves at multiple wavelengths.
In case of RR Lyrae stars, the amplitude and Fourier parameters from the models are consistent with observations in most
period bins except for low metal-abundances ($Z<0.004$). In case of Cepheid variables, we observe a greater offset between
models and observations for both the amplitude and Fourier parameters. The theoretical amplitude parameters are typically
larger than those from observations, except close to the period of $10$ days. We find that these discrepancies between
models and observations can be reduced if a higher convective efficiency is adopted in the pulsation codes. Our results
suggest that a quantitative comparison of light curve structure is very useful to provide constraints for the input physics
to the stellar pulsation models.
}
\vspace{0.5cm}
\section{Introduction}

Classical Cepheids and RR Lyraes are well-known distance indicators, thanks to their period-luminosity relation (Leavitt \& Pickering 1912) that is often used for extragalactic distance measurements and to estimate a precise value of the Hubble constant (Riess et al. 2016). These variables are also very useful stellar tracers of ages and metallicities, and very sensitive probes for the understanding of the theory of stellar pulsation and evolution (Cox 1980).  Simon \& Lee (1981) were the first to use Fourier decomposition method for studying the light curve structure of Cepheids and since then, this method has been widely used for quantitative analysis of the light curves of these variables (Petersen 1984; Jurcsik \& Kovacs 1996; Smolec 2005). In the past decade, several attempts have been made to compare the pulsation properties of Cepheid and RR Lyrae with models (Bono et al. 2000a; Marconi et al. 2013, 2015, 2017). More recently, an extensive comparison of the modern observed light curve data from the OGLE survey (Soszy\'{n}ski et al. 2008, 2010, 2014, 2016, 2017) with the theoretical models has been carried out for Cepheids by Bhardwaj et al. (2017) and for RR Lyraes by Das et al. (2018).

In this paper, we summarize the results on the variation of theoretical light curve parameters of Cepheids and RR Lyraes and their comparison with the observations. The structure of this paper is as follows: Section 2 outlines the theoretical and observed data used in this analysis and briefly describes the Fourier decomposition method. We present the comparison of different theoretical parameters of Cepheid and RR Lyrae stars with those from observations in Section 3 and finally summarise our results in section 4.

\section{Data \& Methodology}

We analyse the theoretical light curves generated using the nonlinear, time-dependent convective hydrodynamical models of a total of 384 FU Cepheids and 274 FU RR Lyraes obtained from Marconi et al. (2013) and Marconi et al. (2015), respectively. The Cepheid and RR Lyrae models were computed by Marconi et al. (2013, 2015) for a fixed chemical composition and an adopted mass-luminosity relation for a range of effective temperatures using the hydrodynamical code developed by Stellingwerf (1982) and updated by Bono \& Stellingwerf (1994) and Bono et al. (1998, 1999). For a detailed discussion concerning the input physics adopted to construct the evolutionary and pulsational models, the interested reader is referred to the papers above and references therein. The Cepheid models have three metal-abundances $-$ $Z=0.02$, $Z=0.008$ and $Z=0.004$ representative of their population in the Galaxy, Large Magellanic Cloud (LMC) and Small Magellanic Cloud (SMC), respectively while the RR Lyrae models have seven different metal-abundances ranging from $Z=0.02$ to $Z=0.0001$. The average metal-abundances of RR Lyrae stars in the Bulge, LMC and SMC was found to be $Z=0.001$, $Z=0.0006$ and $Z=0.0003$, respectively (see, Das et al. 2018, for details). Each chemical composition has a few sets of stellar masses and luminosities. For the Cepheid models, the canonical mass-luminosity relations are adopted from the stellar evolutionary calculations of Bono et al. (2000b) for a fixed chemical-composition. On the other hand, the non-canonical relations adopt a luminosity level brighter by $0.25$ dex corresponding to each mass to account for possible mass-loss and/or overshooting. The convective efficiency is similar in both sets of models. The predicted bolometric light curves of both the Cepheid and RR Lyrae models have been transformed into visual and infrared color curves using static model atmospheres (Castelli et al. 1997a,b). The RR Lyrae models also include models with periods longer than $1$ day to take into account the possibility of evolved RR Lyrae stars.

For a comparison with the models, we analyse the observed light curve data for RR Lyrae stars in the Galactic bulge, LMC and SMC from the OGLE-IV survey (Soszy\'{n}ski et al. 2014, 2016) in the optical ($VI$) bands and in the globular cluster M4 (NGC 6121) from Neeley et al. (2015) in the mid-infrared (3.6 $\mu$m and 4.5 $\mu$m) bands. We also include the photometric data of type II Cepheids in the Bulge (Soszy\'{n}ski et al. 2017), LMC (Soszy\'{n}ski et al. 2008) and SMC (Soszy\'{n}ski et al. 2010) for a comparison with the longer period RR Lyrae models. The photometric data for Cepheids are compiled from literature as given in Bhardwaj et al. (2015). In addition, we also analyse the infrared data for SMC Cepheids from VMC survey (Ripepi et al. 2016, Scowcroft et al. 2016).

The theoretical and photometric light curve data of RR Lyrae and Cepheid variables are fitted with a Fourier sine series of the form (see, Deb \& Singh 2009; Bhardwaj et al. 2017; Das et al. 2018, for more details):

\begin{equation}
m(x) = m_0 + \sum_{k=1}^{N}A_k sin(2 \pi kx+\phi_k),
\label{eq:fourier}
\end{equation}

\noindent where $x$ is the pulsation phase, $m_0$ is the mean magnitude and $N$ is the order of the fit obtained from the Bart's criterion (Bart 1982). Fourier amplitude and phase coefficients ($A_k$ and $\phi_k$) are redefined as Fourier amplitude and phase parameters, respectively:
\begin{equation}
R_{k1} = \frac{A_k}{A_1}; \phi_{k1} = \phi_k - k\phi_1,
\label{eq:params}
\end{equation}
\noindent where, $k > 1$ and $\rm 0 \leq \phi_{k1} \leq 2\pi$.

\section{Light curve analysis of Cepheids and RR Lyraes}

\begin{figure}[h]
\begin{minipage}{8cm}
\centering
\includegraphics[width=8.5cm]{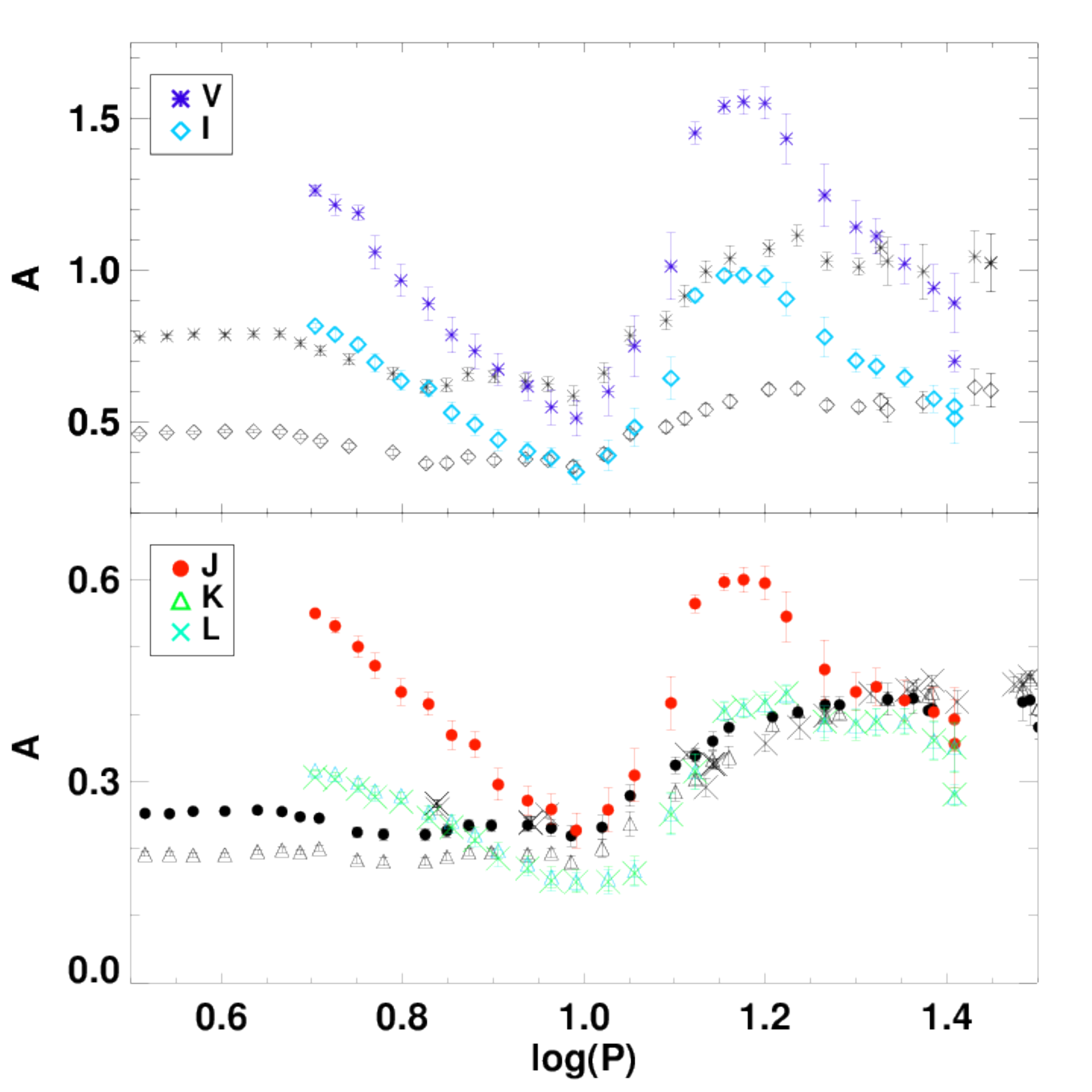}
\caption{A comparison of the mean amplitudes from theoretical (colored) and observed (black) FU Cepheids in the LMC over multiple wavelengths. Taken from Bhardwaj et al. (2017).\label{amp_cep}}
\end{minipage}
\hfill
\begin{minipage}{8cm}
\centering
\includegraphics[width=8.5cm]{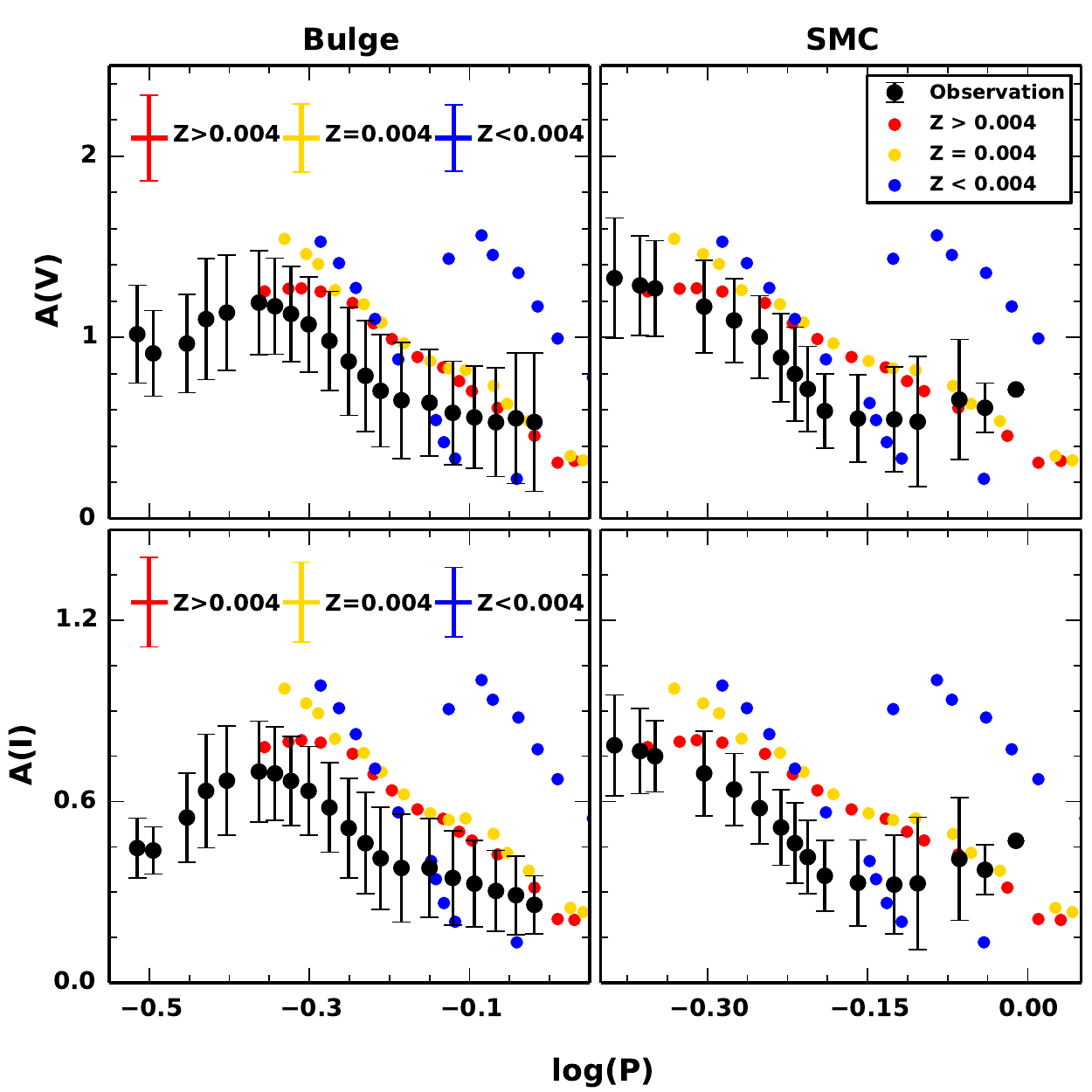}
\caption{A comparison of the mean amplitudes from RRab models (colored) and observations (black) for RRab stars in Bulge and SMC in $V$ and $I$-bands. \label{amp_rrl}}
\end{minipage}
\end{figure}

We compare peak-to-peak amplitudes from the light curves of Cepheid and RR Lyrae variables at multiple wavelengths with those from the models. Fig.\,\ref{amp_cep} shows the comparison of mean amplitudes at optical and near-infrared wavelengths for Cepheids while Fig.\,\ref{amp_rrl} displays the same for RR Lyrae at optical wavelengths. The mean amplitudes are obtained by taking a bin size of $\log(P) = 0.1$ dex, moving in steps of $0.03$ dex and finding the average in the given bin, with the error bars representing the standard deviation on the mean in that particular bin. In case of Cepheids, LMC models with $Z=0.008$ are overplotted in colored symbols in Fig.\,\ref{amp_cep}. We find that the $K$ and $L$-band theoretical amplitudes are consistent with those from observations in most period bins. Similarly, optical ($VI$) and $J$-band amplitudes are also consistent around a period of $10$ days but they are systematically larger than the observed amplitudes in other period bins. The observed mean amplitudes range from $0.1$ to $1.2$ mag while the theoretical amplitudes range upto $\sim 1.6$ mag across infrared and optical bands. We also study the comparison of the mean amplitudes from RRab models and observations in the Bulge and SMC in the optical ($VI$) bands in Fig.\,\ref{amp_rrl} and find a monotonic decrease in the mean amplitudes with period for both models with $Z>0.004$ and $Z=0.004$ and observations in the range $-0.35<\log(P)<0$. While the amplitude ranges are consistent between models and observations, the theoretical mean amplitudes are, in general, higher than those from observations; the longer period ($\log(P)>-0.1$) models with $Z<0.004$ have significantly larger amplitudes than those from observations. An increase in the mixing length parameter would reduce this discrepancy in the results between models and observations by decreasing the pulsation amplitudes of the theoretical light curves, while keeping the structure of the light curves unchanged for both Cepheids (Bono et al. 2002, Bhardwaj et al. 2017) and RR Lyraes (Di Criscienzo et al. 2004).

\begin{figure}[h]
\centering
\includegraphics[scale=1]{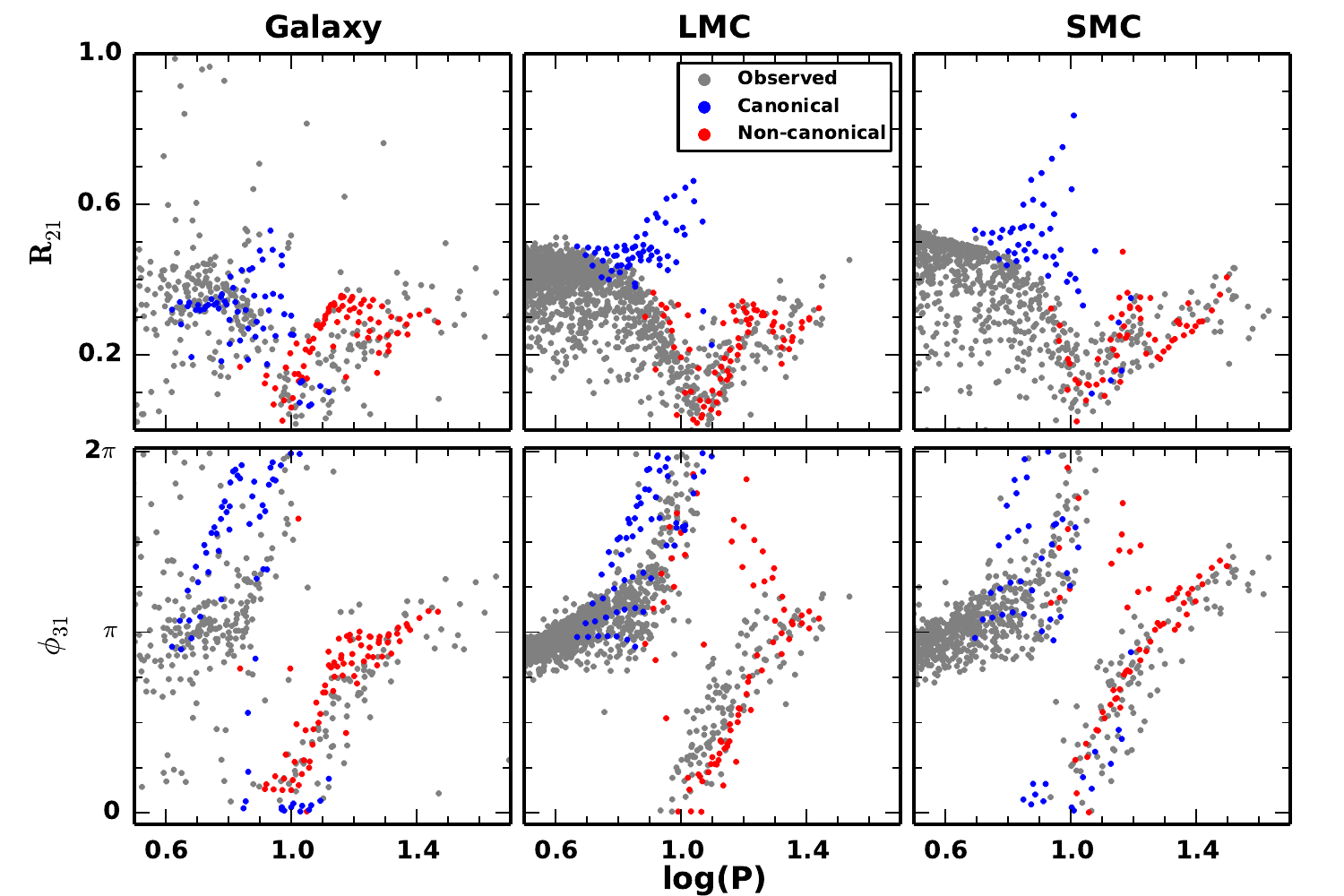}
\caption{A comparison of the $I$-band Fourier parameters of Cepheids from models (colored) and observations (grey) in the Galaxy, LMC and SMC. The non-canonical models (red) have luminosities higher than the canonical models (blue) of the same mass by $0.25$ dex.\label{fourier_cep}}
\end{figure}

\begin{figure}[h]
\centering
\includegraphics[scale=0.8]{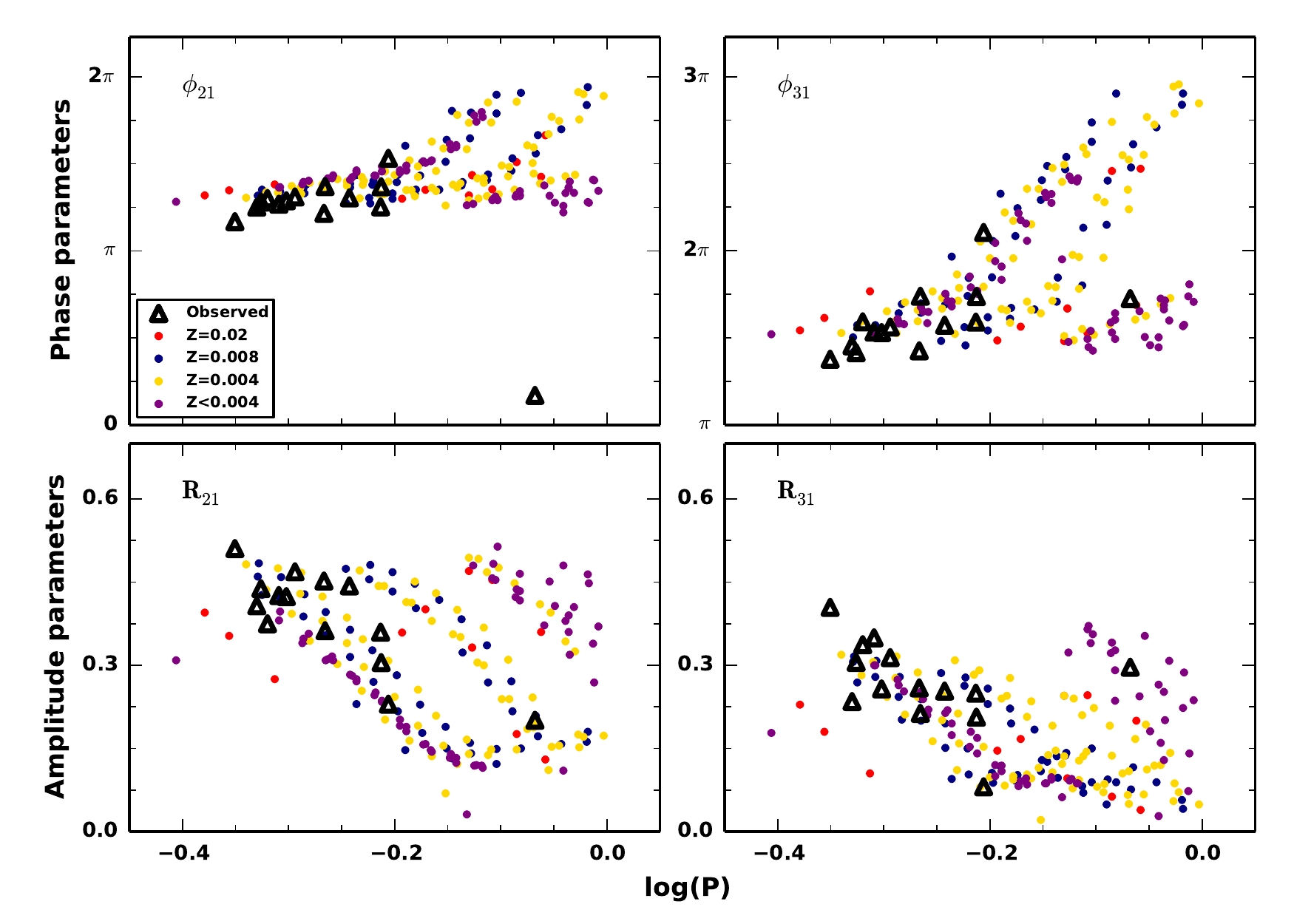}
\caption{A comparison of the theoretical Fourier parameters for RRab with different compositions in $K$-band with observed Fourier parameters from the globular cluster M4 RRab stars in 3.6 $\mu$m band.\label{fourier_rrl_nir}}
\end{figure}

Fig.\,\ref{fourier_cep} to Fig.\,\ref{fourier_rrl_nir} display the comparison of Fourier parameters from models and observations as a function of period for both Cepheid and RR Lyrae variables. Fig.\,\ref{fourier_cep} shows the comparison of Fourier parameters from the observed Cepheids in the Galaxy, LMC and SMC with those from models of corresponding metal-abundances of $Z=0.02$, $Z=0.008$ and $Z=0.004$, respectively in the $I$-band. The theoretical $R_{21}$ are consistent with those obtained from the observed light curves of Cepheids in the Galaxy at the short period range ($\log(P)< 1.1$) and for Cepheids in the Magellanic Clouds at longer periods ($\log(P)>1.1$). However, the amplitude parameters are systematically larger for LMC and SMC for $\log(P)<1$. In the overlapping period range ($0.8 <\log(P)< 1.1$), the canonical models are inconsistent with observations in the $R_{21}$ plane while the non-canonical models match well; this discrepancy between models and observations increases with a decrease in metallicity in the given period range. The theoretical $\phi_{31}$ values have a small offset with the observations from LMC and SMC at the longer periods ($\log(P)>1.2$) but are significantly larger for the short period ($\log(P)<0.9$) Galactic Cepheids.

\begin{figure}[h]
\centering
\includegraphics[scale=1]{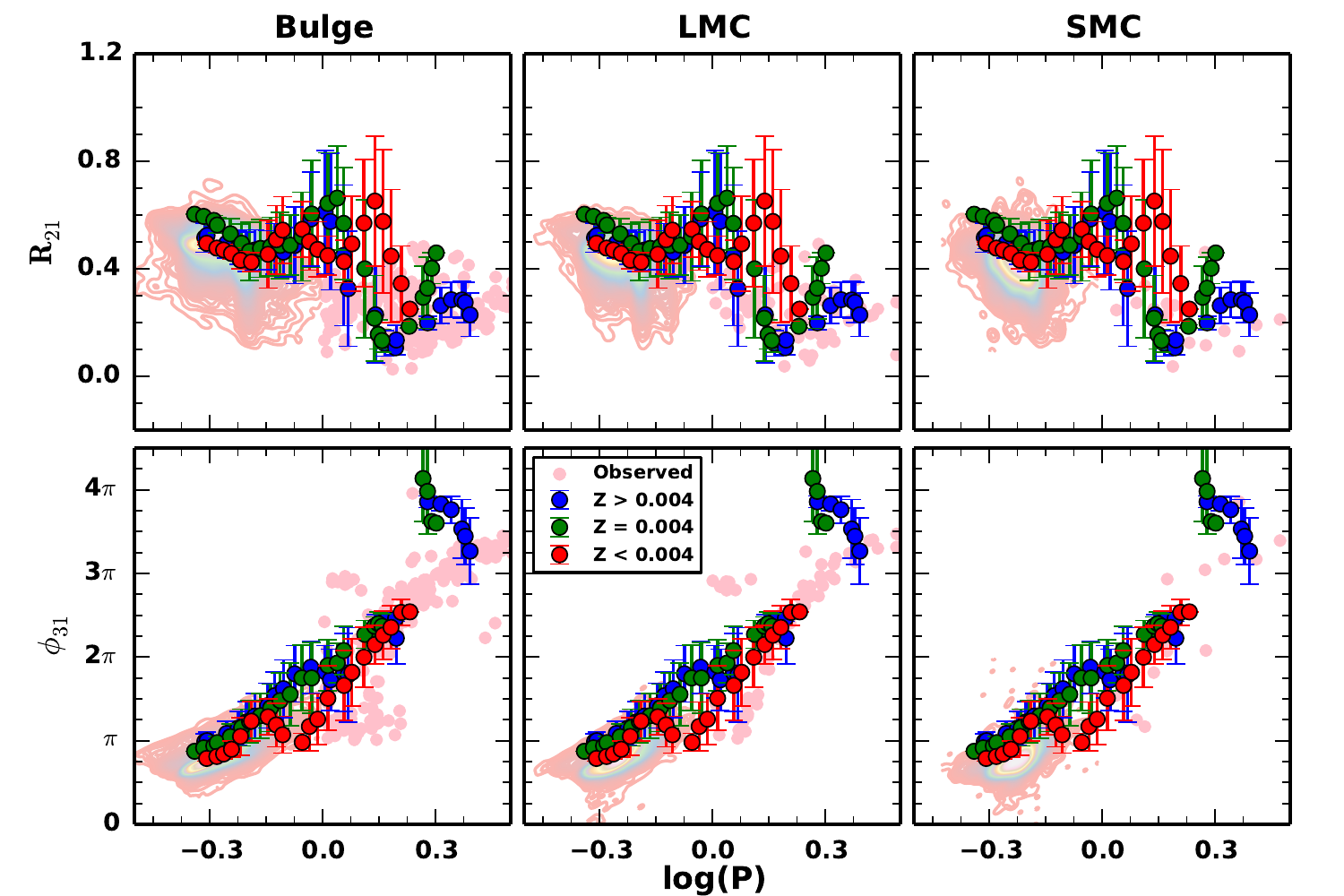}
\caption{A comparison of the mean $I$-band Fourier parameters from RRab models (colored) and observations (pink contour lines for RRab stars and pink scatter points for Type II Cepheids) in the Bulge, LMC and SMC.\label{fourier_rrl}}
\end{figure}

Fig.\,\ref{fourier_rrl} displays a comparison of the mean $I$-band Fourier parameters from RRab models with those from observed RRab stars and type II Cepheids in the Bulge, LMC and SMC. We find the theoretical $R_{21}$ and $\phi_{31}$ values are consistent in most period bins except in the period range of $0<\log(P)<0.15$ where the $R_{21}$ from the $Z<0.004$ are significantly higher than the observed values for the type II Cepheids. While there is an overall consistency in Fourier parameters between models and observations given the large uncertainties, the amplitude parameters are
typically larger than the observations and this discrepancy can be resolved by adopting a higher mixing length (Marconi \& Clementini 2005). We also compared Fourier parameters from limited NIR data available in the globular cluster in Fig.\,\ref{fourier_rrl_nir} and found the observations and theory to be in a better agreement at longer wavelengths. More well-sampled NIR light curves of RR Lyrae are needed to provide more constraints for the theoretical models.

\begin{figure}[h]
\centering
\includegraphics[scale=0.35]{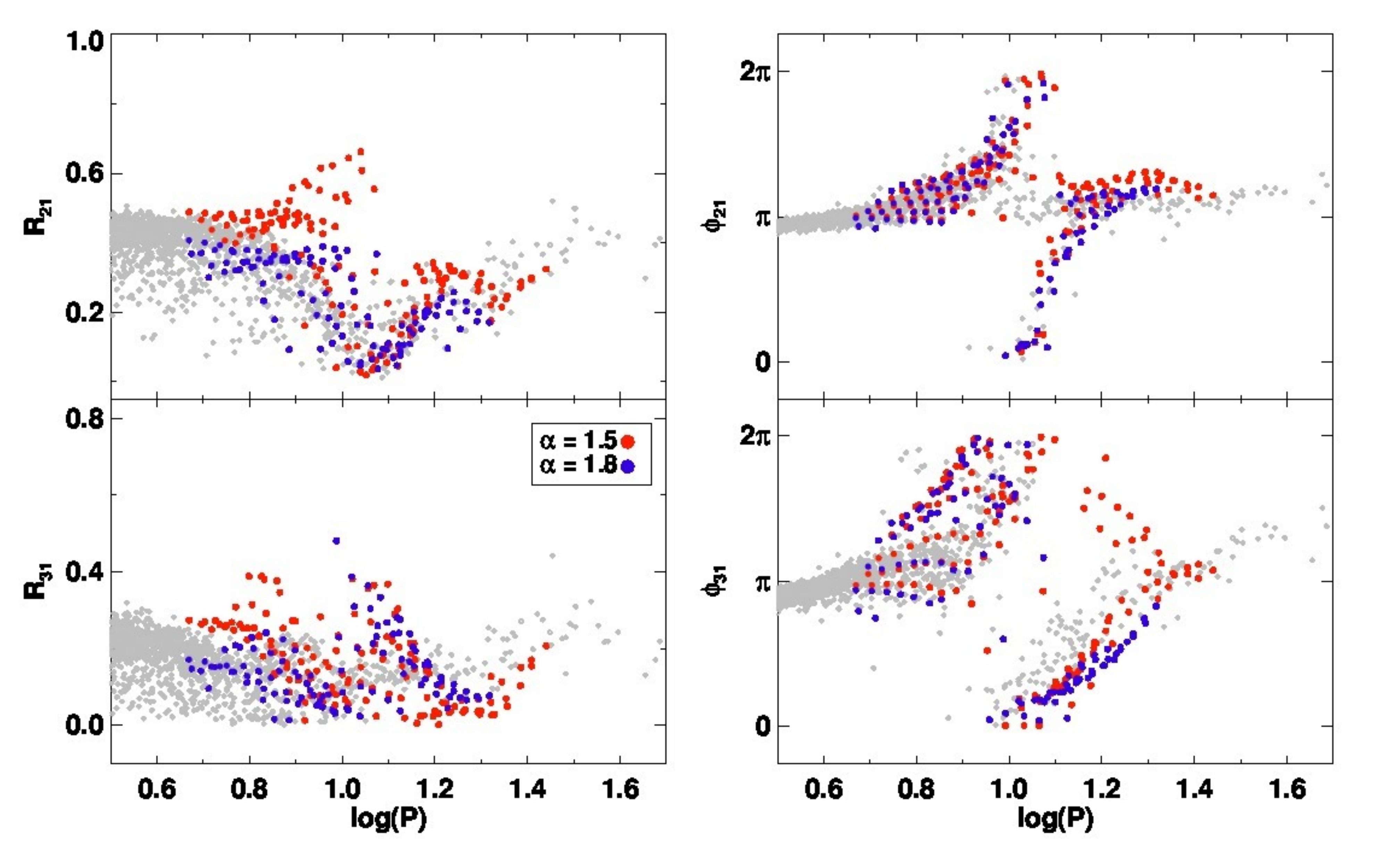}
\caption{A comparison of the $I$-band Fourier parameters for the FU Cepheid models with $Z=0.008$ as a function of different mixing length parameters. The observed parameters from OGLE LMC Cepheids are shown in grey.\label{mixinglength}}
\end{figure}

It has been shown previously that the discrepancies between Fourier parameters from models and observations can be resolved by increasing the mixing length parameter (Marconi \& Clementini 2005, Marconi \& Degl'Innocenti 2007 for RR Lyraes and Fiorentino et al. 2007 for classical Cepheids). Fig.\,\ref{mixinglength} shows the $I$-band Fourier parameters for the FU Cepheid models with $Z=0.008$ for two mixing length parameters, $\alpha=1.5$ and $\alpha=1.8$. It is evident that with the higher mixing length the theoretical amplitudes become more consistent with the observations. In terms of phase parameters, there is only marginal changes in the $\phi_{31}$ values, especially at the shorter period range ($\log(P)<1$). In addition to providing a comparison between models and observations, the variation in the light curve parameters with period and wavelength can also provide insights to non-linearity and metallicity effects on the period-luminosity relations of Cepheid and RR Lyrae variables (Bhardwaj et al. 2017).

\section{Discussion and Conclusions}

We have carried out an extensive comparison of the multiband observed light curves of Cepheid and RR Lyrae variables with the most recent nonlinear, time-dependent convective hydrodynamical models. We find that the theoretical amplitude parameters from Cepheids are systematically larger than those from observations as a function of period, except around the period of $10$ days for the $VIJ$-bands. The $KL$-band theoretical amplitudes are more consistent with observations over most period bins. For the RR Lyrae models, barring the longer period ($\log(P)>-0.1$) models with $Z<0.004$, the theoretical amplitude parameters are well-consistent with those from observations. However, the theoretical peak-to-peak amplitude parameters from both Cepheids and RR Lyraes are, in general, higher from those from observations. Increasing the mixing length parameter would cause a decrease in the pulsation amplitudes of the theoretical light curves for both Cepheids (Bono et al. 2002) and RR Lyraes (Di Criscienzo et al. 2004), thereby reducing the offset between the models and observations. These results form a basis for a rigorous comparison between pulsation models and observations for Cepheid and RR Lyrae and will provide stringent constraints for the pulsation models when combined with upcoming data from the multiwavelength wide-field variability surveys. As a part of our future work, we plan to observe RR Lyrae stars in the Globular clusters at multiple wavelengths to investigate amplitude and phase modulations in these variables and to constrain the metallicity dependence on their period-luminosity relations at near-infrared wavelengths. Additionally, the bright classical Cepheids and long-period variables in the M31/M33 will be monitored with one or more BINA telescopes including the 3.6-m Devasthal Optical Telescope for the stellar pulsation and distance scale studies.

\section*{Acknowledgements}
The authors thank the referee for useful comments and suggestions. HPS acknowledges support from the Belgo-India Network in Astronomy (BINA) project and India PI Dr. Santosh Joshi and kind hospitality of the Royal Observatory of Belgium, Brussels. HPS and SMK thank the Indo-US Science and Technology Forum for funding the Indo-US virtual joint networked centre on ``Theoretical analyses of variable star light curves in the era of large surveys''. SD acknowledges the INSPIRE Junior Research Fellowship vide Sanction Order No. DST/INSPIRE Fellowship/2016/IF160068 under the INSPIRE Program from the Department of Science \& Technology, Government of India. AB acknowledges the research grant \#11850410434, awarded by the National Natural Science Foundation of China through a Research Fund for International Young Scientists and China post-doctoral general grant. AB, SMK and MM acknowledge support by the Munich Institute for Astro- and Particle Physics (MIAPP) of the DFG cluster of excellence ``Origin and Structure of the universe'' during their stay at the MIAPP workshop on ``Extragalactic Distance Scale in the Gaia Era''.

%
%
%

\footnotesize
\beginrefer

\refer Bart M. L. 1982, IJNA, 2, 241

\refer Bhardwaj A., Kanbur S. M., Singh H. P., Macri L. M., Ngeow C.-C. 2015, MNRAS, 447, 3342

\refer Bhardwaj A., Kanbur S. M., Marconi M. et al. 2017, MNRAS, 466, 2805

\refer Bono G., Stellingwerf R.F. 1994, ApJS, 93, 233

\refer Bono G., Caputo F., Marconi M. 1998, ApJ, 497, L43

\refer Bono G., Marconi M., Stellingwerf R.F. 1999, ApJS, 122, 167

\refer Bono G., Castellani V., Marconi M. 2000a, ApJL, 532, L129

\refer Bono G., Caputo F., Cassisi S. et al. 2000b, ApJ, 543, 955

\refer Bono G., Castellani V., Marconi M. 2002, ApJ, 565, L83

\refer Castelli F., Gratton R. G., Kurucz R. L. 1997a, A\&A, 318, 841

\refer Castelli F., Gratton R. G., Kurucz R. L. 1997b, A\&A, 324, 432

\refer Cox J. P. 1980, Theory of stellar pulsation

\refer Das S., Bhardwaj A., Kanbur S. M., Singh H. P., Marconi M. 2018, MNRAS, 481, 2000

\refer Deb S., Singh H. P. 2009, A\&A, 507, 1729

\refer Di Criscienzo M., Marconi M., Caputo F. 2004, MmSAI, 75, 190

\refer Fiorentino G., Marconi M., Musella I., Caputo F. 2007, A\&A, 476, 863

\refer Jurcsik J., Kovacs G. 1996, A\&A, 312, 111

\refer Leavitt H. S., Pickering E. C. 1912, HarCi, 173, 1

\refer Marconi M., Clementini G. 2005, AJ, 129, 2257

\refer Marconi M., Degl'Innocenti S. 2007, A\&A, 474, 557

\refer Marconi M., Molinaro R., Bono G. et al. 2013, ApJ, 768, L6

\refer Marconi M., Coppola G., Bono G. et al. 2015, ApJ, 808, 50

\refer Marconi M., Molinaro R., Ripepi V. et al. 2017, MNRAS, 466, 3206

\refer Neeley J. R., Marengo M., Bono G. et al. 2015, ApJ, 808, 11

\refer Petersen J. O. 1984, A\&A, 139, 496

\refer Riess A. G., Macri L. M., Hoffmann S. L. et al. 2016, ApJ, 826, 56

\refer Ripepi V., Marconi M., Moretti M. I. et al. 2016, ApJs, 224, 21

\refer Scowcroft V., Freedman W. L., Madore B. F. et al. 2016, ApJ, 816, 49

\refer Simon N. R., Lee A. S. 1981, ApJ, 248, 291

\refer Smolec R. 2005, AcA, 55, 59

\refer Soszy\'nski I., Udalski A., Szyma\'nski M. K. et al. 2008, AcA, 58, 293

\refer Soszy\'nski I., Udalski A., Szyma\'nski M. K. et al. 2010, AcA, 60, 91

\refer Soszy\'nski I., Udalski A., Szyma\'nski M. K. et al. 2014, AcA, 64, 177

\refer Soszy\'nski I., Udalski A., Szyma\'nski M. K. et al. 2016, AcA, 66, 131

\refer Soszy\'nski I., Udalski A., Szyma\'nski M. K. et al. 2017, AcA, 67, 297

\refer Stellingwerf R. F. 1982, ApJ, 262, 330

\endrefer           

\end{document}